\RequirePackage{lineno}
\documentclass[twocolumn,letterpaper,aps,prc,longbibliography,superscriptaddress,nofootinbib,floatfix]{revtex4-1}

\usepackage{graphicx}	
\usepackage{appendix}
\usepackage{float}
\usepackage{bbm}
\usepackage{hyperref}
\usepackage{amsmath}
\usepackage{verbatim}
\usepackage{xspace}

\hypersetup{
    colorlinks=true,
    linkcolor=blue,
    filecolor=blue,
    urlcolor=blue,
    citecolor=blue,
}

\usepackage{xspace}	

\newcommand{\pt}{\mbox{$p_T$}\xspace}

\newcommand{\nch}{\mbox{$N_{\rm ch}$}\xspace}
\newcommand{\dnchdeta}{\mbox{$dN_{\rm ch}/d\eta$}\xspace}

\newcommand{\sqsn}{\mbox{$\sqrt{s_{_{NN}}}$}\xspace}

\newcommand{\pp}{\mbox{$p$$+$$p$}\xspace}
\newcommand{\dau}{\mbox{$d$$+$Au}\xspace}

\newcommand{\pau}{\mbox{$p$$+$Au}\xspace}
\newcommand{\pa}{\mbox{$p$$+$$A$}\xspace}

\newcommand{\ppb}{\mbox{$p$$+$Pb}\xspace}
\newcommand{\po}{\mbox{$p$$+$O}\xspace}
\newcommand{\pbpb}{\mbox{Pb$+$Pb}\xspace}
\newcommand{\nucnuc}{\mbox{$A$$+$$A$}\xspace}
\newcommand{\xexe}{\mbox{Xe$+$Xe}\xspace}
\newcommand{\oo}{\mbox{O$+$O}\xspace}

\newcommand{\heau}{\mbox{$^{3}$He$+$Au}\xspace}

\newcommand{\pcc}{\mbox{$\rho(v^{2}_{n},[p_{T}])$}\xspace}

\newcommand{\pythia}{\mbox{{\sc pythia8}}\xspace}
\newcommand{\angantyr}{\mbox{{\sc pythia-angantyr}}\xspace}
\newcommand{\ampt}{\mbox{{\sc ampt}}\xspace}
\newcommand{\hijing}{\mbox{{\sc hijing}}\xspace}

\newcommand{\ipglasma}{\mbox{{\sc ip-glasma}}\xspace}
\newcommand{\ipjazma}{\mbox{{\sc ip-jazma}}\xspace}

\begin{document}

\title{Exploring Origins for Correlations between Flow Harmonics and Transverse Momentum in Small Collision Systems\\
(Unambiguous Ambiguity)
}

\newcommand{\pusan}{Pusan National University, Busan, 46241, South Korea}

\newcommand{\colorado}{University of Colorado, Boulder, Colorado 80309, USA}
\affiliation{\colorado}

\author{S.H.~Lim} \affiliation{\pusan}
\author{J.L.~Nagle} \affiliation{\colorado}

\date{\today}

\begin{abstract}
High statistics data sets from experiments at the Relativistic Heavy Ion  Collider (RHIC) and the Large Hadron Collider (LHC) with small and large collision species have enabled a wealth of new flow measurements, including the event-by-event correlation between observables.    One exciting such observable $\rho(v^{2}_{n},[p_{T}])$ gauges the correlation between the mean transverse momentum of particles in an event and the various flow coefficients ($v_n$) in the same event~\cite{PhysRevC.93.044908}.   Recently it has been proposed that very low multiplicity events may be sensitive to initial-state glasma correlations~\cite{Giacalone:2020byk} rather than flow-related dynamics. 
We find utilizing the \ipjazma framework that the color domain explanation for the glasma results are incomplete.
We then explore predictions from \pythia, and the version for including nuclear collisions called \angantyr, which have only non-flow correlations and the \ampt model which has both non-flow and flow-type correlations.   
We find that \angantyr has non-flow contributions to $\rho(v^{2}_{n},[p_{T}])$ in \po, \ppb, \oo collisions that are positive at low multiplicity and comparable to the glasma correlations.   It is striking that in \pythia in \pp collisions there is actually a sign-change from positive to negative $\rho(v^{2}_{n},[p_{T}])$ as a function of multiplicity.
The \ampt results match the experimental data general trends in Pb+Pb collisions at the LHC, except at low multiplicity where \ampt has the opposite sign.   In \ppb collisions, \ampt has the opposite sign from experimental data and we explore this within the context of parton geometry.
Predictions for \po, \oo, and \xexe are also presented.
\end{abstract}


\maketitle



\section{Introduction}
\label{sec:intro}

The standard time-evolution model for relativistic heavy ion collisions at the Relativistic Heavy Ion Collider (RHIC) and the Large Hadron Collider (LHC) includes an ``epoch'' of quark-gluon plasma (QGP) formation and hydrodynamic expansion~\cite{Busza:2018rrf}.    The QGP expands as a strongly coupled fluid with very small specific shear viscosity $\eta/s$~\cite{Romatschke:2017ejr,Heinz:2013th}.  Detailed computer modeling of said time  evolution requires inputs including the initial energy deposit geometry and any pre-hydrodynamic evolution.    Higher level ``flow'' correlation observables, such as the event-by-event correlation of anisotropies ($v_{n}^{2}$) and  mean transverse momentum $([p_{T}])$, can provide additional insights and constraints on the interplay of geometry and subsequent transport~\cite{PhysRevC.93.044908}.   

The ATLAS experiment has carried out measurements of the $\rho(v^{2}_{n},[p_{T}])$ defined as \begin{align}
    \rho(v^{2}_{n},[p_{T}]) = \frac{{\rm cov}(v_{n}^{2},[p_T])}{\sqrt{{\rm Var}(v_{n}^{2})} \sqrt{{\rm Var}([p_T])}}
\end{align}
in \ppb and  \pbpb collisions at  \sqsn = 5.02~TeV at the LHC~\cite{Aad:2019fgl}. The ${\rm cov}(v_{n}^{2},[p_T])$ is the covariance between $v_n^{2}$ and $[p_T]$, and ${\rm Var}(v_{n}^{2})$ and ${\rm Var}([p_T])$ are the variances of $v_n^{2}$ and $[p_T]$. Focusing on the correlation of elliptic flow $v_{2}^{2}$ and $[p_T]$, in \pbpb collisions the correlator is negative at low event multiplicity, then turning positive and increasing with multiplicity, and finally slightly decreasing for  the very highest multiplicity events.    These features are semi-quantitatively reproduce in a simple calculation of initial Glauber geometry combined with hydrodynamic  flow ~\cite{Bozek:2016yoj,Giacalone:2020dln,Schenke:2020uqq}, though the results depend sensitively  on the geometry model in small collision systems~\cite{Bozek:2020drh}.

The basic idea is that there are event-to-event fluctuations in the geometric size of the overlap region in the transverse plane at fixed multiplicity.   Smaller overlap area with the same multiplicity leads to larger pressure gradients and thus larger radial flow and hence larger mean transverse momentum $\left< p_{T} \right>$.  If there is a positive correlation in geometry for smaller overlap area events to have a larger eccentricity, and vice versa, then one gets a positive $\rho$ correlator.   If there is the opposite correlation such that for smaller overlap area events, they have a smaller eccentricity, and vice versa, then one gets a negative $\rho$ correlator.   It has been shown that alternative variables to the overlap area present a stronger correlation~\cite{Giacalone:2020dln,Schenke:2020uqq}; however, the basic picture above still holds.

However, the detailed behaviour in the lowest  multiplicity  events warrants further examination.   The question of whether small collision systems, for example \pp and \ppb at the LHC and \pau, \dau, \heau at RHIC, have correlations induced via hydrodynamic expansion of small QGP droplets or via a mimic from initial-state correlations, for example glasma correlations, has been studied in detail~\cite{Nagle:2018nvi,Dusling:2015gta}.   Extensive data from the LHC and the geometry scan from  RHIC~\cite{PHENIX:2018lia}, combined with theoretical clarification for glasma correlations~\cite{Mace:2018vwq,Mace:2018yvl,Nagle:2018ybc}, have given a clear answer for the higher multiplicity events in small collision systems -- there is overwhelming evidence for correlations via initial geometry coupled  with strong final-state interactions (e.g. hydrodynamics).  

Even more recently, it has been proposed that initial-state correlations may yet dominate over final-state collectivity in the very lowest multiplicity  collisions, i.e. for $\dnchdeta < 10$~\cite{Schenke:2019pmk}.   Current there  is no clear experimental evidence for such glasma correlations at these multiplicities.   A proposal has been put forth that the correlation between anisotropies $v_{n}^{2}$ and event average $p_T$ can provide definitive evidence:    
Ref.~\cite{Giacalone:2020byk} states that ``experimental observation of these
clean qualitative signatures in peripheral heavy ion and small system collisions will be the first evidence for the presence and importance of initial state momentum anisotropies predicted by an effective theory of QCD.''

The calculations, in a hybrid glasma and hydrodynamic model, indicate that the $\rho$ correlator should change sign from negative at low multiplicity back to positive at the very lowest multiplicities $\dnchdeta < 10$ where glasma correlations dominate.    The basic  argument is that in the glasma color domain picture, events with a smaller transverse size will have fewer color domains and hence a stronger glasma type correlation.   Ref.~\cite{Schenke:2021mxx} notes that at fixed multiplicity, the saturation scale $Q_{s}$ of the  projectile, which drives the $\langle \pt \rangle$, decreases with increasing transverse area and hence the positive $\rho$ correlator from initial-state glasma correlations alone.
We highlight  that the connection between the calculation in momentum-space and the spatial domain picture (see Ref.~\cite{Lappi_2016} for details) makes the cause-effect relation for the positive $\rho$ correlation via glasma diagrams to some degree speculative.   In  Appendix~\ref{AppendixB}, we utilize the \ipjazma framework to show that the scaling of projectile saturation scale and eccentricity  with overlap area do  not follow the ordering postulated above and thus require further validation.

Due diligence requires the checking of whether non-exotic (i.e. not glasma  correlations) could also mimic said sign change and hence positive $\rho$ correlations at the lowest multiplicities.    
It is notable that within the calculation of Ref.~\cite{Giacalone:2020byk}, there is no inclusion of what are often referred to as non-flow correlation effects (e.g. from dijet correlations, overall momentum conservation, etc.).   


In this paper we explore the expectations for the $\rho$ correlator in the \pythia and \angantyr models~\cite{Sjostrand:2007gs,Bierlich_2018}, which have neither initial-state glasma correlations nor final-state interactions.   \angantyr is an extension of the publicly available \pythia code that allows the modeling of \pa and \nucnuc collisions, and is named for Angantyr the Berserker in Norse mythology~\cite{berserker}.
We highlight that there is a version of \angantyr with so-called ``string shoving''~\cite{Bierlich:2016vgw} that may effectively model final-state interactions; however, we are running the code without this option turned on.  

We also explore calculations from the A-Multi-Phase Transport (\ampt) model~\cite{Lin_2005}, which again has no initial-state glasma correlations but has both initial state non-flow (e.g. dijets) and final-state parton and hadron scattering.   
In the process of completing this manuscript, a complementary study using the \pythia and \hijing~\cite{Gyulassy_1994} models has been submitted~\cite{zhang2021nonflow}.


\section{Methodology}

There are different methods for constructing the $\rho$ correlator, in part in an effort to reduce non-flow contributions. We follow the three-subevent method~\cite{Aad:2019fgl} using three separate $\eta$ regions labeled $a$, $b$, and $c$: $-2.5< \eta_{a}<-0.75$, $|\eta_{b}|<0.5$, and $0.75<\eta_{c}<2.5$.
Charged particles in subevents $a$ and $c$ are used to calculate $v_{n}^{2}$ to suppress the non-flow contribution, and the $[p_{T}]$ is obtained with charged particles from subevent $b$.
The \pcc is calculated using particles in $0.3<p_T<2.0~\mathrm{GeV}$, and particles in $|\eta|<2.5$ and $0.5<p_T<5.0~\mathrm{GeV}$ are used to calculate the event multiplicity \nch.

The ${\rm cov}(v_{n}^{2},[p_T])$ term in the $\rho$ correlator is defined as 
\begin{align}
    {\rm cov}(v_{n}^{2},[p_T]) = {\rm Re} \left( \left< \sum_{a,c} e^{in(\phi_{a} - \phi_{c})} \left( [p_T] - \langle [p_T] \rangle \right) \right> \right),
\end{align}
where $\phi_{a(c)}$ is the azimuthal angle of particles in the region $a$ ($c$), and $\langle [p_T] \rangle$ is the average $[p_T]$ in the entire events at a certain multiplicity range.
The variance of $v_{n}^{2}$ is calculated with two-particle and four-particle correlations,
\begin{align}
    {\rm Var}(v_{n}^{2}) &= v_{n}\{2\}^{4} - v_{n}\{4\}^{4} \nonumber\\
    &= \langle {\rm corr}_{n}\{4\} \rangle - \langle {\rm corr}_{n}\{2\} \rangle^{2},
\end{align}
where $v_{n}\{2\}$ and $v_{n}\{4\}$ are the flow coefficients obtained from two- and four-particle correlations, respectively. 
The ${\rm corr}_{n}\{2\}$ and ${\rm corr}_{n}\{4\}$ are the two-particle and four-particle correlations with the subevent method~\cite{Jia:2017hbm} with particles in the region $a$ and $c$.
Events required to have at least two charged particles in each region ($a$ and $c$) for the four-particle correlation.  
The variance of $[p_T]$ is estimated by the dynamical \pt fluctuation magnitude~\cite{Abelev:2014ckr} $c_k$ defined as
\begin{align}
    c_k = \left< \frac{1}{N_{\rm pair}} \sum_{b}\sum_{b^{\prime}\neq b} (p_{T,b} - \langle [p_T] \rangle )(p_{T,b^{\prime}} - \langle [p_T] \rangle) \right>, 
\end{align}
and particles in region $b$ are used. 
The \pcc is calculated with ${\rm cov}(v_{n}^{2},[p_T])$, ${\rm Var}(v_{n}^{2})$, and $c_k$, and the ${\rm Var}(v_{n}^{2})$ should be positive for valid \pcc value. 


\section{\pythia Results}

\begin{figure}
    \centering
    \includegraphics[width=0.9\linewidth]{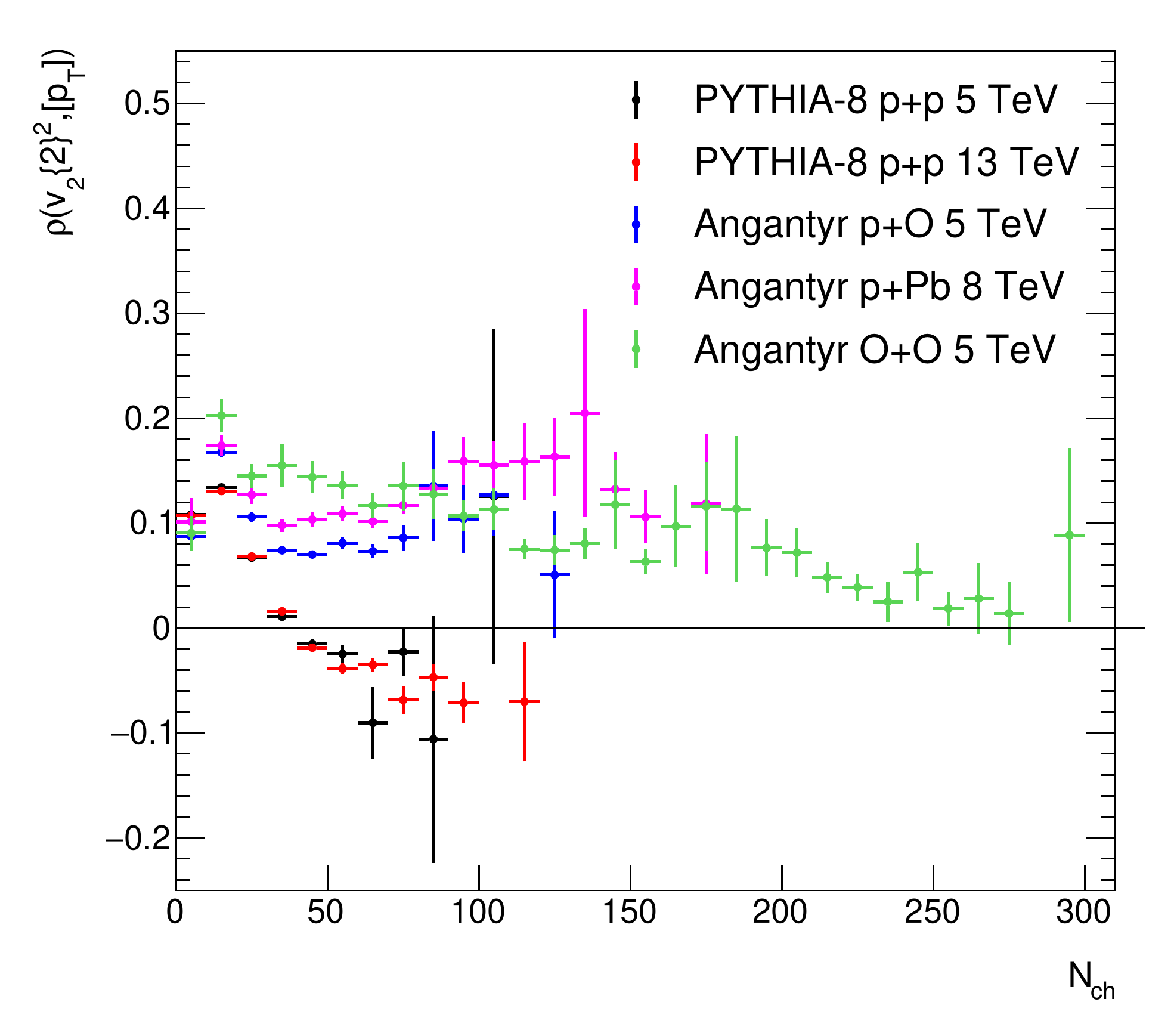}
    \caption{Results shown from  the \pythia model for \pp at 5 and 13 TeV and the \angantyr model for \po, \oo at 5 TeV, \ppb at 8 TeV.
    In  all cases the event categories on the $x$-axis are defined by \nch for charged hadrons with $|\eta|<2.5$ and $0.5 < \pt < 5.0~\mathrm{GeV}$.  The particles entering the $\rho$ correlations are from $0.3 < \pt <2.0~\mathrm{GeV}$ and are utilized using the three-subevent method.}
    \label{fig:pythia}
\end{figure}

We have run millions of events for various collision systems within the \pythia and \angantyr frameworks.    The output particle truth list is then processed through the full three-subevent calculation detailed above.   Here we utilize the ATLAS define regions of $-2.5< \eta_{a}<-0.75$, $|\eta_{b}|<0.5$, and $0.75<\eta_{c}<2.5$ and consider charged hadrons with $0.3 < p_{T} < 2.0~\mathrm{GeV}$.   For the event category, the \nch is determined within the range $-2.5 < \eta < 2.5$ and for charged hadrons with $0.5 < p_{T} < 5.0~\mathrm{GeV}$.  Figure~\ref{fig:pythia} shows the resulting $\rho$ correlator 
from \pythia for \pp at 5 TeV and 13 TeV and from \angantyr for \po at 5 TeV, \ppb at 8 TeV, and \oo at 5 TeV.    Note that future running of \po and \oo at the LHC may end up being run at a different collision energy.    The full eight-panel set of calculated ingredients that go into the $\rho$ correlator are shown in the Appendix~\ref{appendixA} in Figure~\ref{fig:pythiasmall8}.

The correlations in \po, \ppb, and \oo collisions are positive for all \nch and are around 10\%.   In contrast, the results in \pp at 5 and 13 TeV (consistent with each other) are positive for $\nch < 30$ and then change sign to negative for $\nch > 30$.    These non-zero correlations cannot be from spatial geometry coupled to final-state interactions or flow, since \angantyr has neither.    Instead they result from non-trivial correlations due to jets, correlation between jets and underlying event, and multiplicity category correlations.   

\begin{figure}
    \centering
    \includegraphics[width=0.9\linewidth]{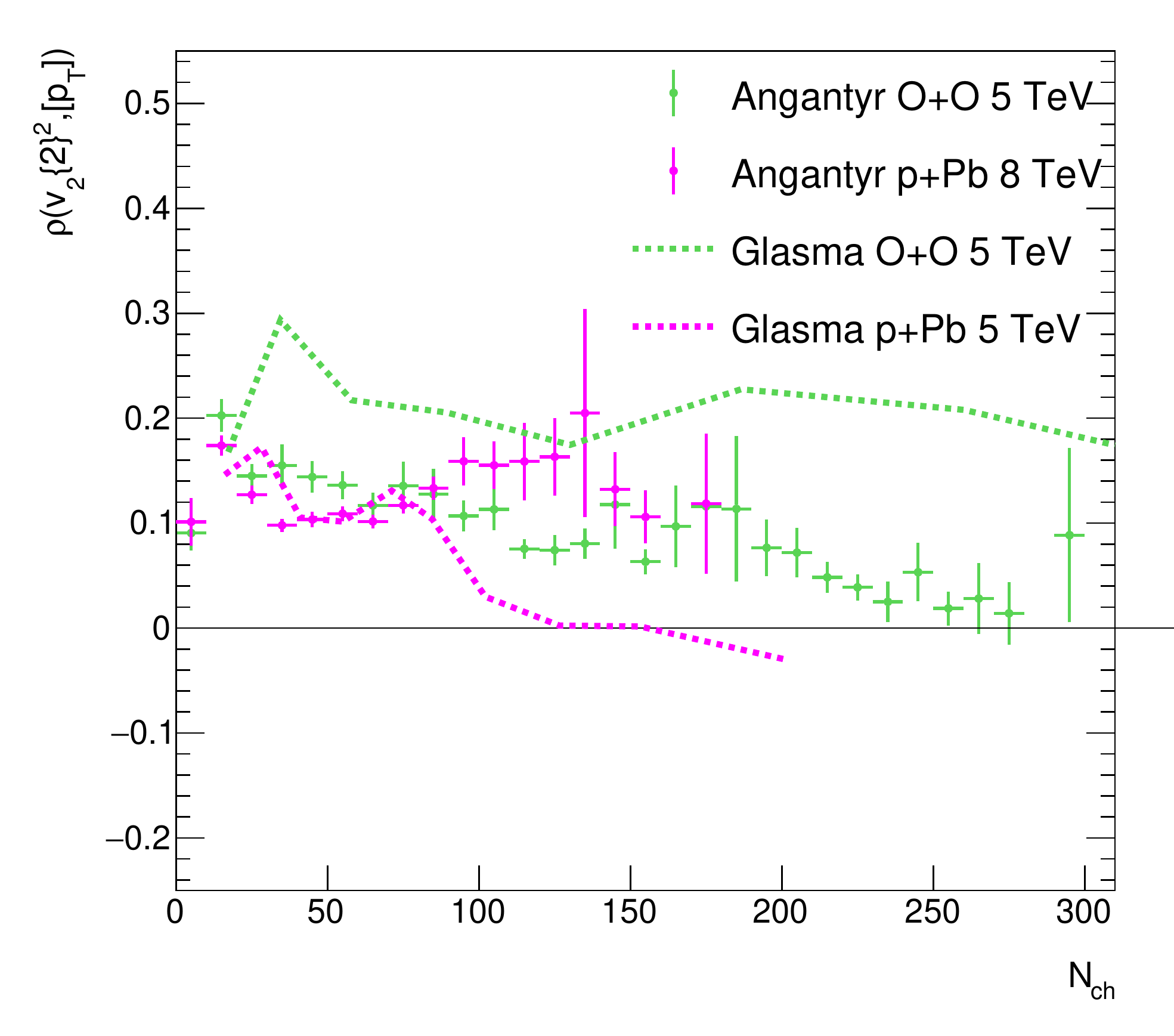}
    \caption{Results shown from the \angantyr model for \ppb at 8 TeV and \oo at 5 TeV.
    In  all cases the event categories on the $x$-axis are defined by \nch for charged hadrons with $|\eta|<2.5$ and $0.5 < \pt < 5.0~\mathrm{GeV}$.  The particles entering the $\rho$ correlations are from $0.3 < \pt <2.0~\mathrm{GeV}$ and are utilized using the three-subevent method.  In comparison are results from initial-state glasma correlations~\cite{Giacalone:2020byk}.}
    \label{fig:pythia-glasma}
\end{figure}

Figure~\ref{fig:pythia-glasma} shows a comparison in \ppb and \oo between the \angantyr results and glasma initial-state calculations~\cite{Giacalone:2020byk}.   In this case, the glasma correlator is a proxy determined in momentum space from the correlations encoded in the $T_{\mu \nu}$ prior to any hydrodynamic calculation in their hybrid framework.   What is striking is that both calculations are positive and of a similar order of magnitude.    It decidedly means that a full accounting of non-flow effects, as coded into \pythia and \angantyr, are needed to make any statement about more exotic glasma correlations.  

It should be be pointed out that such contributions as seen in the $\rho$ correlation may be additive to leading order if the sources of flow and non-flow particles are independent, as is often assumed in non-flow subtraction techniques.   However, in low multiplicity events this independence must be violated due to momentum conservation, overall color neutrality, as well as the possible direct scattering of particles from each source.    Thus, one can postulate additive contributions~\cite{zhang2021nonflow}, though at the lowest multiplicities it is unknown the level of violation.

\section{\ampt Results}

We now compare with calculations from the \ampt model (version v1.26t9b-v2.26t9b),  which is publicly available code~\cite{Lin_2005}.    The \ampt model has been run it the default mode, with a parton screening mass of 3.22~fm$^{-1}$ (or equivalently a parton-parton cross section of 3~mb) and with Lund symmetric splitting function parameters ($a=0.4$ and $b=0.8$), set to optimize matching with particle multiplicities and $\left< p_{T} \right>$ at the LHC~\cite{Zhang:2019utb}.  \ampt has been successful at describing a number of global and flow features of heavy ion data, including in small collision systems -- see Refs.~\cite{Zhang:2019utb,He_2016,Koop:2015wea,Bzdak_2014} for examples.

\begin{figure}
    \centering
    \includegraphics[width=0.9\linewidth]{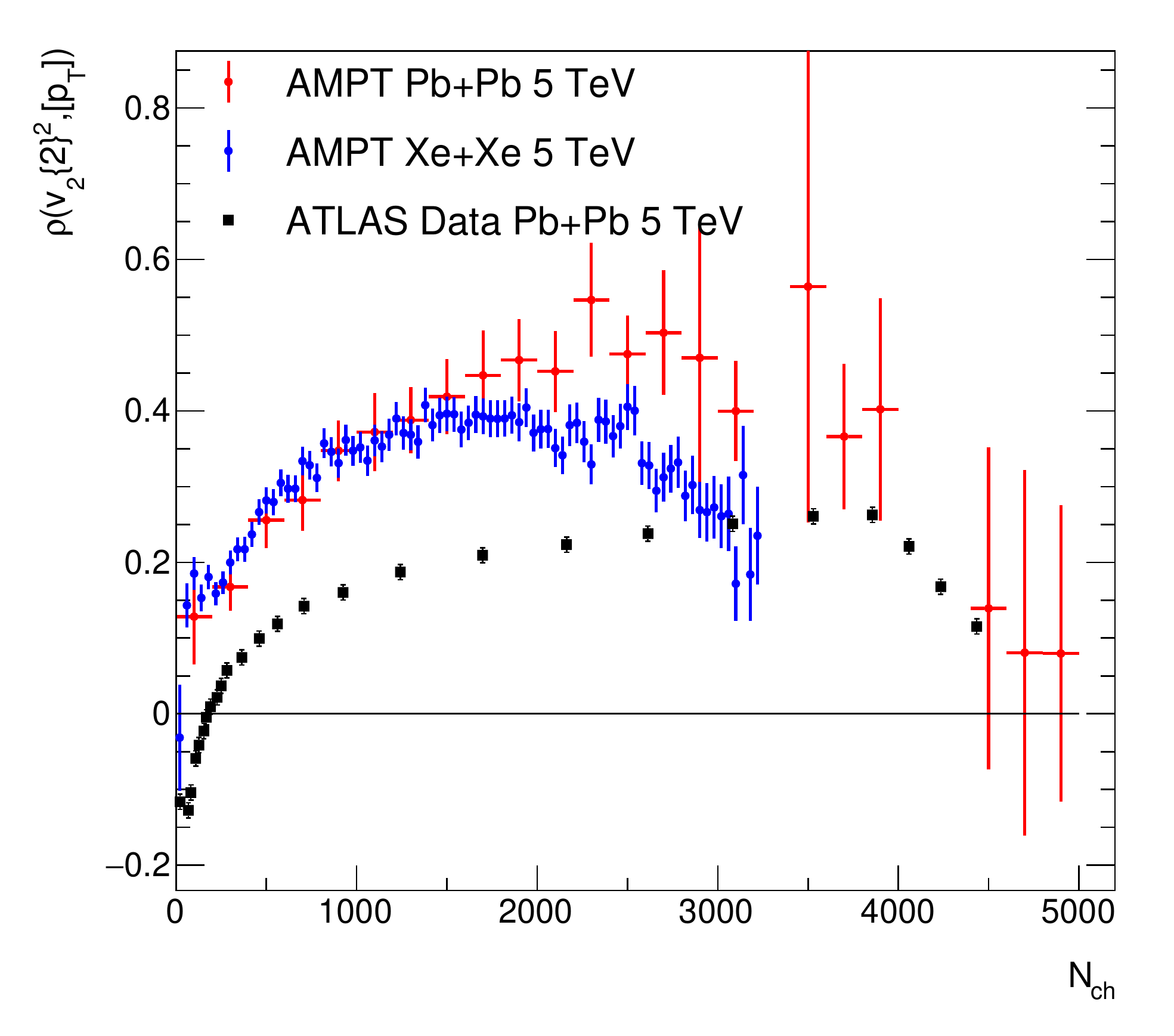}
    \caption{Results shown from  the \ampt model for \pbpb and \xexe at 5 TeV compared with published ATLAS results for \pbpb collisions at the same energy.
    In  all cases the event categories on the $x$-axis are defined by \nch for charged hadrons with $|\eta|<2.5$ and $0.5 < \pt < 5.0~\mathrm{GeV}$.   The particles entering the $\rho$ correlations are from $0.5 < \pt <2.0~\mathrm{GeV}$ and are utilized using the three-subevent method.}
    \label{fig:ampt_pbpb}
\end{figure}

Again, millions of \ampt events have been run for \pp at 5 and 13 TeV, \po and \oo at 5 TeV, \ppb at 5 and 8 TeV, and \xexe and \pbpb at 5 TeV.   The results for large \nucnuc systems are shown in Figure~\ref{fig:ampt_pbpb} and compared with ATLAS results in \pbpb collisions~\cite{Aad:2019fgl}.   The AMPT results are calculated with the three-subevent method in $-2.5< \eta_{a}<-0.75$, $|\eta_{b}|<0.5$, and $0.75<\eta_{c}<2.5$ and consider charged hadrons with $0.5 < p_{T} < 2.0~\mathrm{GeV}$, matching the higher low-$p_{T}$ selection used by ATLAS in Ref.~\cite{Aad:2019fgl}.  
The ATLAS data $x$-axis \nch values are re-scaled since ATLAS does not correction for reconstruction efficiency in the published results.    

The \ampt \pbpb results show a qualitatively similar trend to the ATLAS data, but are approximately a factor of two higher.    It is also notable that neither the \ampt \pbpb nor \xexe go negative for the lowest \nch as seen in the data.    ATLAS has shown a modest dependence in \pbpb collisions of $\rho$ on the $p_{T}$ selection of the charged hadrons used in their analysis~\cite{Aad:2019fgl}.   They observe a modest 15\% increase in  $\rho$ when  switching to  charged hadrons with  $0.5 < \pt < 5.0~\mathrm{GeV}$, while in \ampt the  has a roughly 15\% decrease in $\rho$ with the same momentum selection change.

The full eight panel comparison of ingredients for \pbpb that go into the $\rho$ calculation are shown in Appendix~\ref{appendixA} in Figure~\ref{fig:amptbig8withdata}.   Again one observes qualitatively similar trends between \ampt and data, but significant quantitative differences.   One other striking observation from  \ampt is that the  average $p_T$ as a function of \nch actually decreases with increasing \nch over a significant range of \nch -- see Figure~\ref{fig:amptbig8withdata} in Appendix~\ref{appendixA}.    This is contrary to results in experimental data~\cite{Abelev:2013bla} where average transverse momentum always increases with increasing \nch in \pp, \ppb and \pbpb collisions.   One key question regarding finite parton-parton scattering pictures is how they build up radial flow and thus increase the $\left< p_T \right>$. This \ampt result indicates that this effect is not modeled fully in \ampt, even for large collision systems.

\begin{figure}
    \centering
    \includegraphics[width=0.9\linewidth]{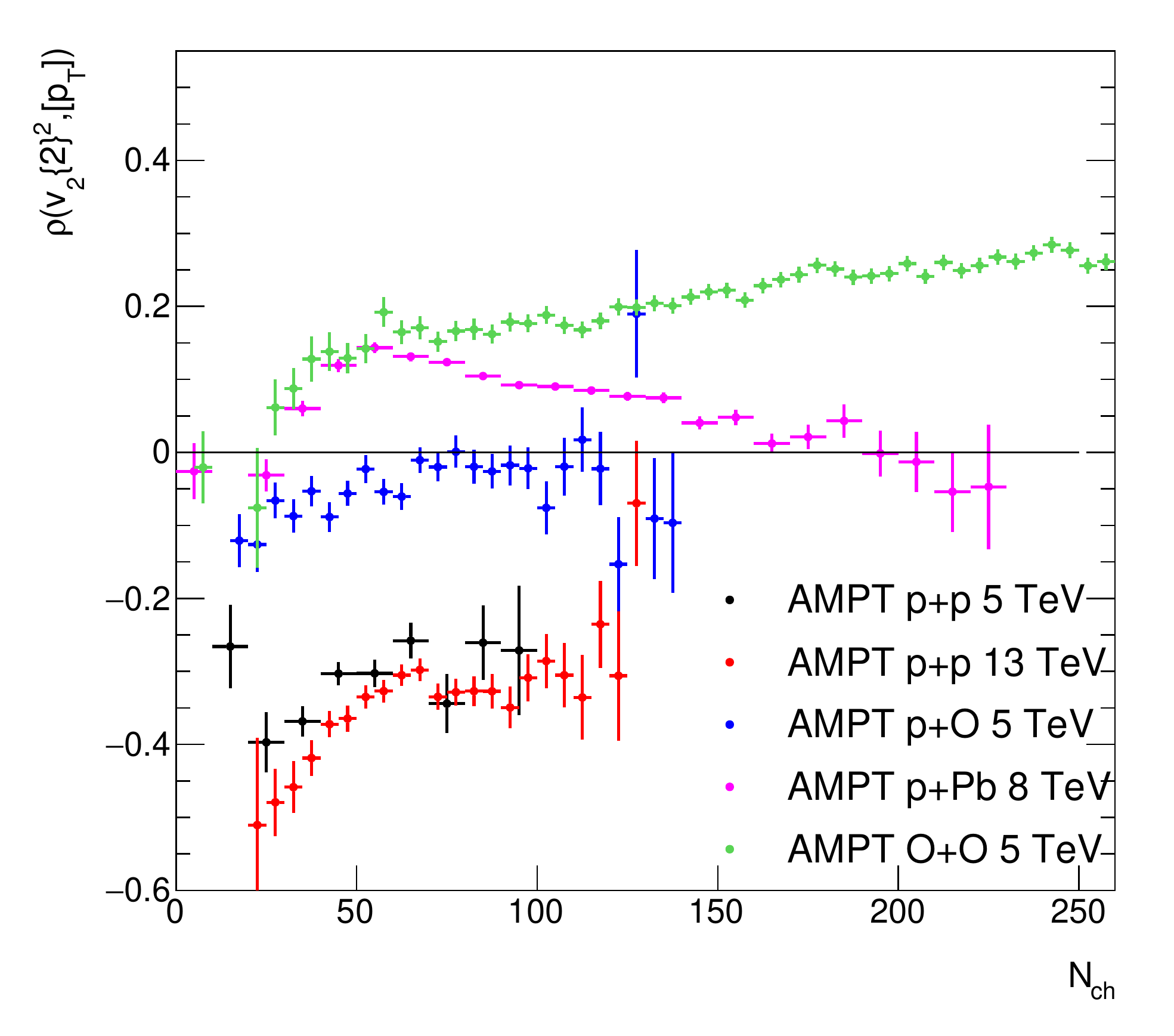}
    \caption{Results shown from  the \ampt model for \pp, \po, \oo at 5 TeV, \ppb at 8 TeV, and \pp at 13 TeV.
    In  all cases the event categories on the $x$-axis are defined by \nch for charged hadrons with $|\eta|<2.5$ and $0.5 < \pt < 5.0~\mathrm{GeV}$.   The particles entering the $\rho$ correlations are from $0.3 < \pt <2.0~\mathrm{GeV}$ and are utilized using the three-subevent method.}
    \label{fig:ampt_small}
\end{figure}

\ampt predictions for small collision systems, \pp at 5 and 13 TeV, \po and \oo at 5 TeV, and  \ppb at 8 TeV are shown in Figure~\ref{fig:ampt_small}.   In this case the results vary from a large negative $\rho$ correlation in \pp collisions to positive correlations in \ppb and \oo.   It  is striking that the positive correlations in \ppb and \oo are quite comparable to those from \angantyr in  Figure~\ref{fig:pythia}, even though \ampt has both flow and non-flow contributions.     

ATLAS has measured the correlator $\rho$ in \ppb collisions, but from earlier data taken at 5 TeV. \ampt calculations are compared with ATLAS data for \ppb collisions at 5 TeV for the selection $0.5 < \pt < 2.0$ in Figure~\ref{fig:ampt_ppb_comp}.    The \ampt result is positive, though decreasing with \nch, and is opposite to the consistently negative $\rho$ result in data.
The full eight panel comparison of ingredients for \ppb that go into the $\rho$ calculation are shown in Appendix~\ref{appendixA} in Figure~\ref{fig:amptsmall8withdata}.

\begin{figure}
    \centering
    \includegraphics[width=0.9\linewidth]{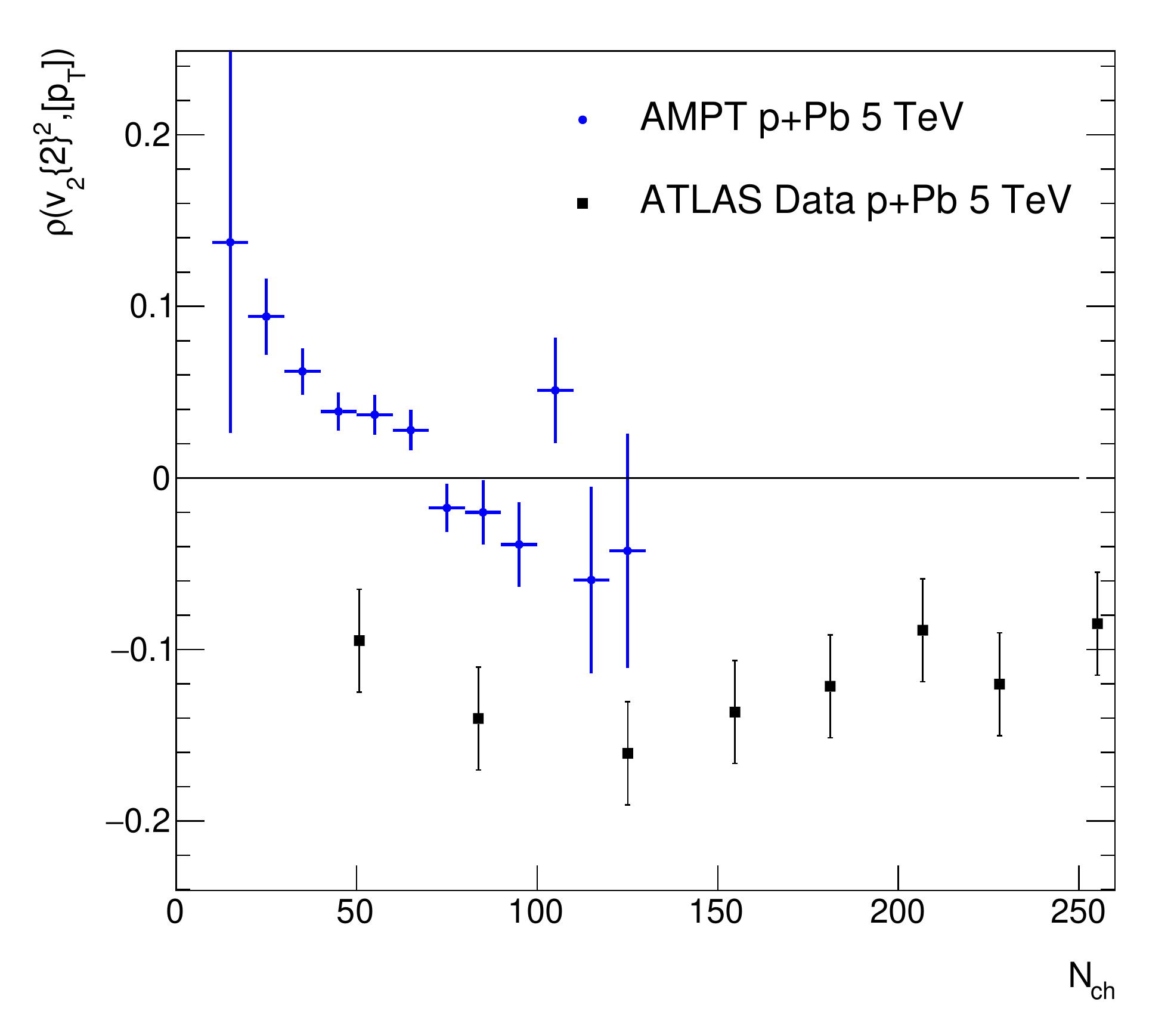}
    \caption{Results shown from  the \ampt model for \ppb at 5 TeV compared with published ATLAS results for the same system and  energy.
    In  all cases the event categories on the $x$-axis are defined by \nch for charged hadrons with $|\eta|<2.5$ and $0.5 < \pt < 5.0~\mathrm{GeV}$.   The particles entering the $\rho$ correlations are from $0.5 < \pt <2.0~\mathrm{GeV}$ and are utilized using the three-subevent method.}
    \label{fig:ampt_ppb}
\end{figure}

\begin{figure}
    \centering
    \includegraphics[width=0.9\linewidth]{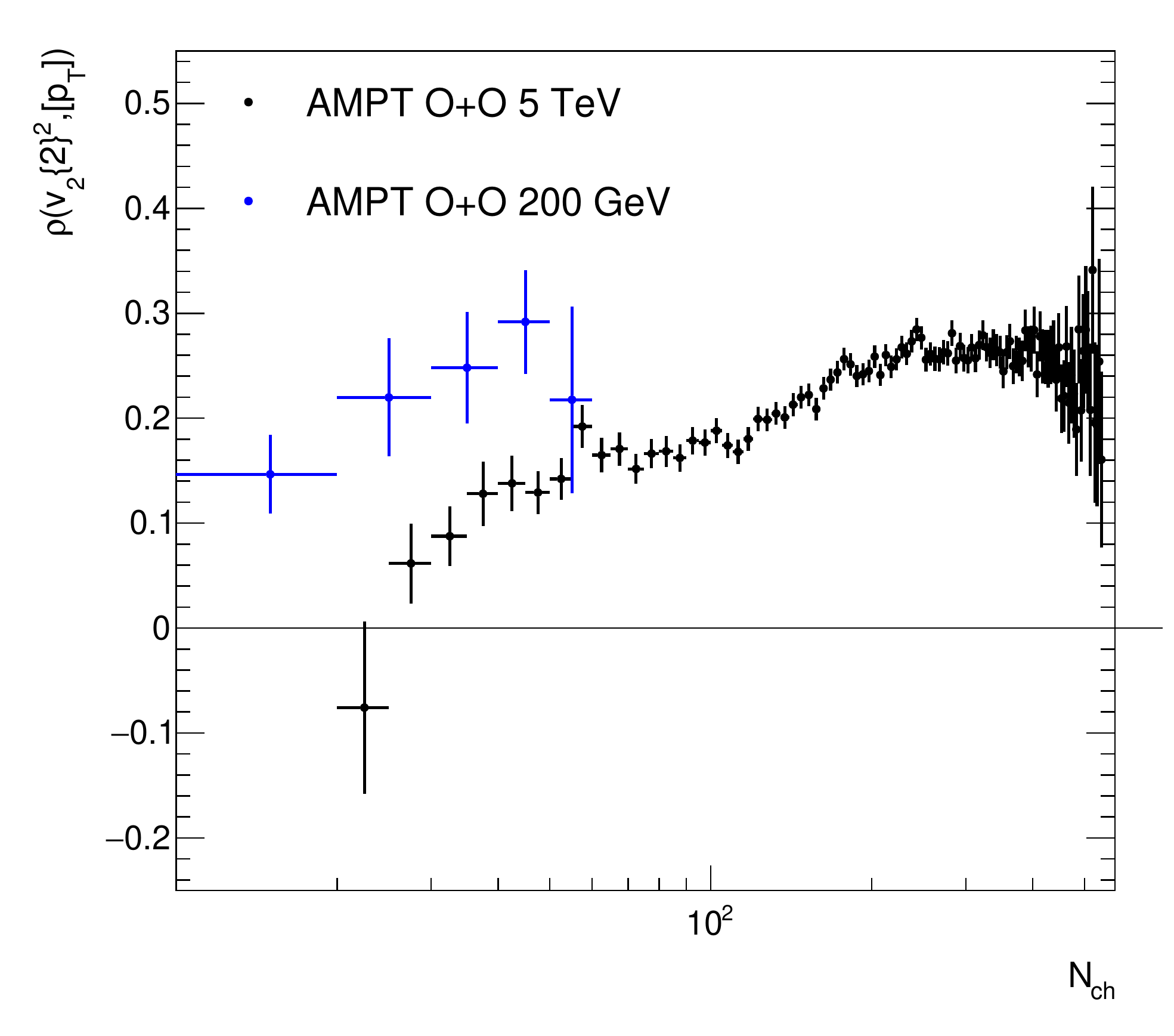}
    \caption{Predictions for \oo at 200 GeV (RHIC) and 5 TeV (LHC) from the \ampt model.  See text for details on the different charged hadron kinematic selections.}
    \label{fig:oo}
\end{figure}

Since there may be \oo data taking at RHIC by the STAR experiment and/or the sPHENIX experiment at 200 GeV, as well as at the LHC in Run-3, we present predictions for the $\rho$ correlator from those systems in Figure~\ref{fig:oo}.   We highlight that for the results at 200 GeV, we have modeled the acceptance selections for a detector with $|\eta|<1.0$ and as such  have regions $-1.0< \eta_{a}<-0.35$, $|\eta_{b}|<0.3$, and $0.35<\eta_{c}<1.0$ and consider charged hadrons with $0.2 < p_{T} < 2.0~\mathrm{GeV}$.   Also, the $x$-axis values of \nch are defined by $|\eta|<0.5$ and $0.2 < p_{T} < 2.0~\mathrm{GeV}$.
The results indicate a positive $\rho$ correlator with a very similar shape and magnitude at the two energies if one considers a rescaling of the $x$-axis between collision energies.

\section{Discussion}

\begin{figure*}[ht]
    \centering
    \includegraphics[width=1.0\linewidth]{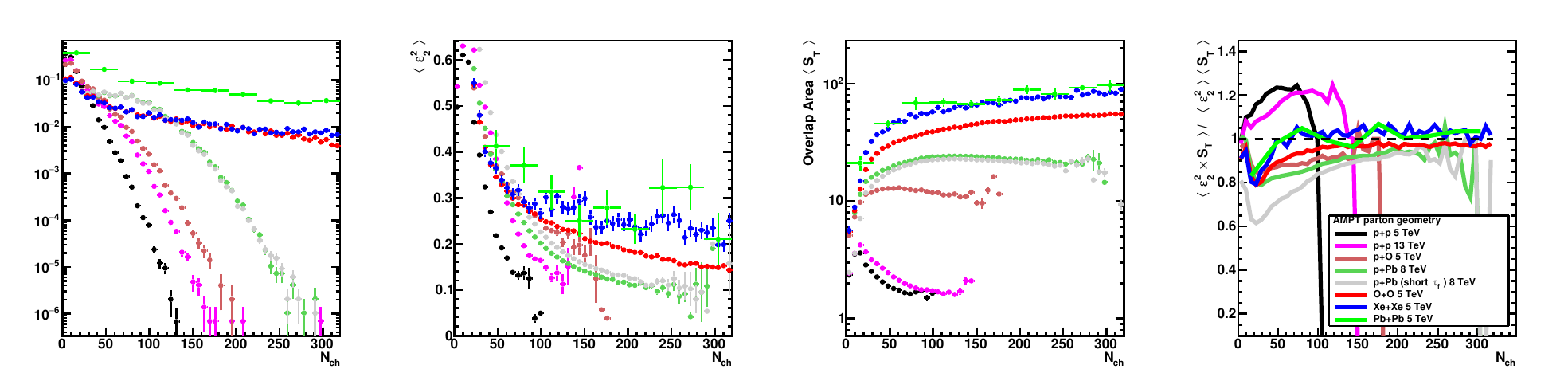}
    \caption{\ampt results for multiple collision systems.   The panels from left to  right show (i) the multiplicity distribution of events normalized to unity, (ii) the average eccentricity squared $\varepsilon_{2}^{2}$ as a function of \nch, (iii) the average overlap area $S_T$ as a function of \nch, and the correlation between $\varepsilon_{2}^{2}$ and $S_T$ normalized by their individual average values.  The \nch axis is zoomed in to focus on the lower multiplicity  range.}
    \label{fig:amptgeom}
\end{figure*}

Within the \ampt model, one can examine the geometry as defined by the partons that are ``string-melted'' out of the color strings -- as has been done previously~\cite{Nagle:2017sjv,Koop:2015trj}.   The partons after appearing have a formation time set in \ampt corresponding to the uncertainty principle before they can start scattering with other partons.    It is their location after the formation time utilized here for calculating the eccentricity of the event $\varepsilon_2$ and the overlap area $S_T$~\cite{Alver:2008zza}.   Here we define the overlap area $S_T = \pi \sqrt{\sigma_{x}^{2}\sigma_{y}^{2}-\sigma_{xy}^{2}}$,  noting there are different prefactors used in the literature.   

Figure~\ref{fig:amptgeom} shows results from \pp at 5 and 13 TeV, \po and \oo at 5 TeV, \ppb at 8 TeV, and \xexe at 5 TeV.   The panels from left to  right show (i) the multiplicity distribution of events normalized to unity, (ii) the average eccentricity squared $\varepsilon_{2}^{2}$ as a function  of \nch, (iii) the average overlap area $S_T$ as a function of \nch, and the (iv)  correlation between $\varepsilon_{2}^{2}$ and $S_T$ normalized by their individual average values.    Thus, the correlation is $> 1$ if there is a correlation between eccentricity and area and $< 1$ if there is an anti-correlation.

The correlation (or anti-correlation) in geometry is a good predictor for \ampt for whether the $\rho$ correlation will be positive or negative.    The positive correlation of eccentricity and area in \pp collisions at 5 and 13 TeV is mirrored in the negative $\rho$ correlation shown in Figure~\ref{fig:ampt_small}, though not at larger \nch.    The negative geometry correlation for \pa and \oo is mirrored in the positive $\rho$ correlation again shown in Figure~\ref{fig:ampt_small}.    

\begin{figure*}
    \centering
    \includegraphics[width=1.0\linewidth]{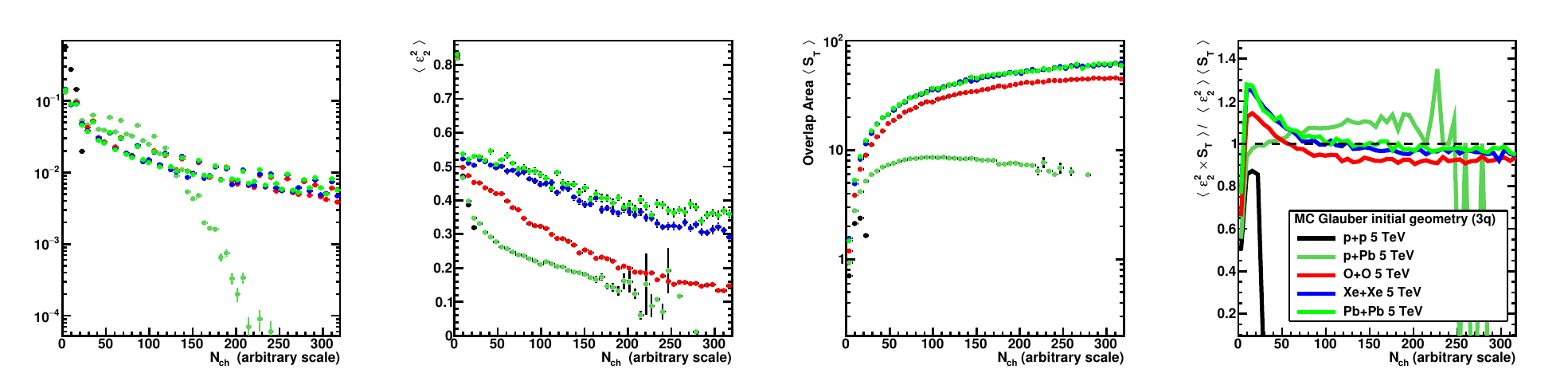}
    \caption{Monte Carlo Glauber  geometries including three constituents for each nucleon.   Note  that the  $x$-axis is simply  an arbitrary scaling times the number of constituent quark-quark collisions.     For \pp collisions in particular this does not model the multiplicity  distribution properly and additional fluctuations are necessary to describe data.
}
    \label{fig:glauber_geom}
\end{figure*}

For comparison, we calculate the geometry correlation for Monte Carlo Glauber type models~\cite{Loizides:2016djv}.   The Glauber  calculation is carried out with standard nuclear geometry parameters and including three constituents per nucleon.
The results for the (i) multiplicity, (ii) eccentricity squared, (iii) the overlap area, and (iv) the correlation of eccentricity and area are shown in Figure~\ref{fig:glauber_geom} in the panels from left to right.

In this zoomed in view for $\nch < 300$, the geometry correlations are often opposite between \ampt and Monte Carlo  Glauber with  three constituents.   We know that in  the case of \pp collisions, the \ampt geometry is almost exclusively two strings and thus the partons have a geometry that appears as two radial distributions of partons around each string with the distance in the transverse plane based on the formation time -- see Ref.~\cite{Nagle:2017sjv} for details.   In  this case, two strings that are farther apart will lead to a much more elliptical geometry and a larger area; hence the positive correlation seen in the  \ampt results in Figure~\ref{fig:amptgeom}.    In contrast, in the Monte Carlo Glauber with three constituents, the larger area typically leads to a more circular geometry and thus a negative correlation. However, in \ampt as soon as one has a \pa collision, there are multiple strings and they have a geometry correlation that is negative and once again flipped from the Monte Carlo Glauber case.    The disagreement of \ampt with experimental \ppb data shown in Figure~\ref{fig:ampt_ppb} may  lead one to conclude that \ampt string geometry is not a good descriptor, or that a substantial over-prediction of non-flow in  \ampt changes the sign.    Additional comparisons with measurements in \pp data should prove useful.

\begin{figure}
    \centering
    \includegraphics[width=0.9\linewidth]{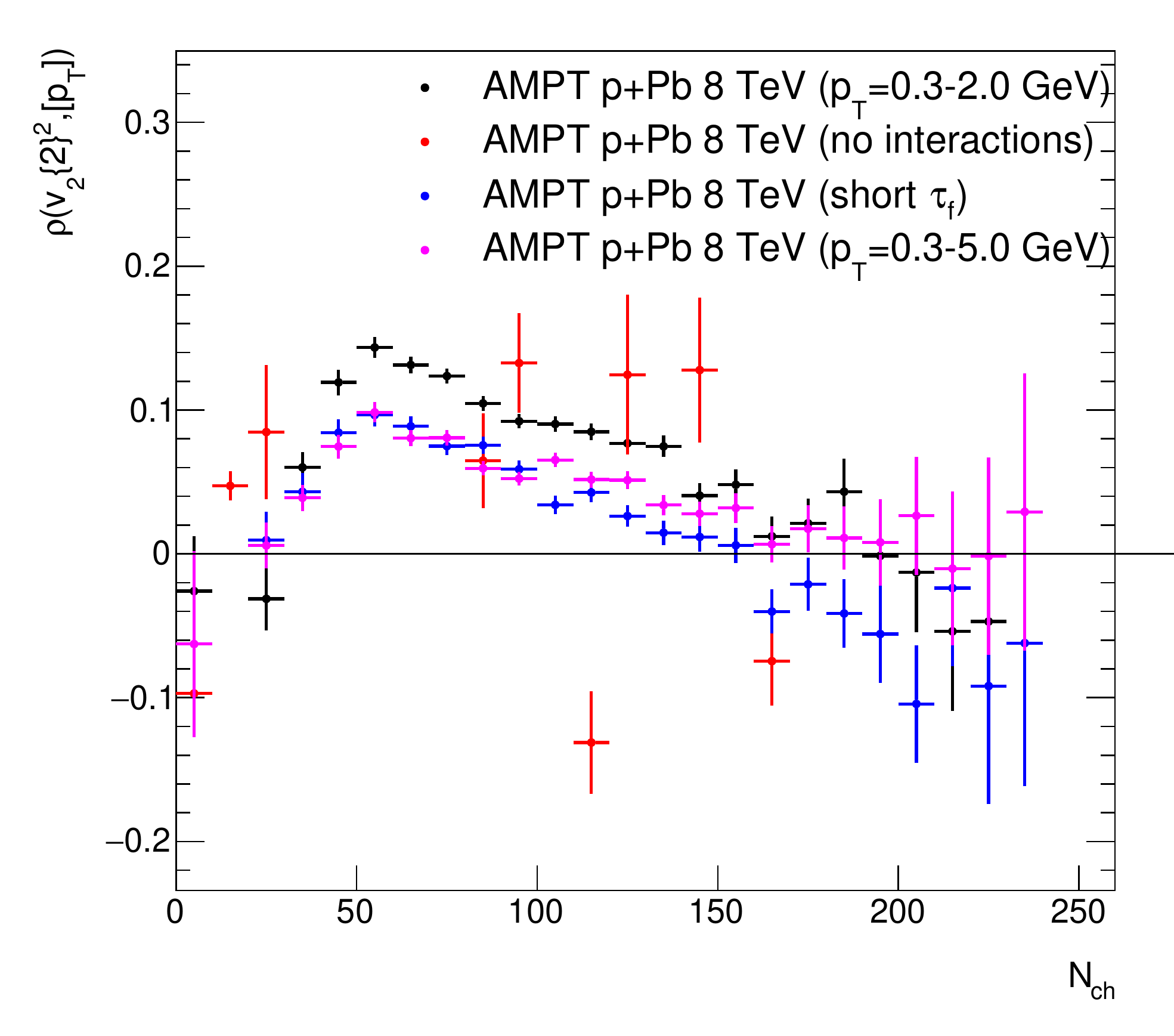}
    \caption{Results shown from  the \ampt model for \ppb at 8 TeV in different running modes.
    In  all cases the event categories on the $x$-axis are defined by \nch for charged hadrons with $|\eta|<2.5$ and $0.3 < \pt < 5.0~\mathrm{GeV}$, except in the comparison case with $0.3 < \pt < 5.0~\mathrm{GeV}$.   Also shown are results with a shorter formation time in \ampt.    Lastly, results with no interactions for \ampt are shown, though they have low statistical significance due to the very small  $v_{2}^{2}$ values overall.}
    \label{fig:ampt_ppb_comp}
\end{figure}

One question is whether the formation time implemented in \ampt might be altering the initial geometry as it mimics free streaming.    Thus, for \ppb collisions, we have re-run \ampt with the formation time reduced by a factor of 10.    Those results, labeled short $\tau_{f}$, are also shown in Figure~\ref{fig:amptgeom} and show a modestly smaller overlap area and a modestly larger eccentricity.   These effects do impact the $\rho$ correlator as shown in Figure~\ref{fig:ampt_ppb_comp}.   These changes do modify the $\rho$ value by almost  a factor of two, though do not change the sign.

\begin{figure}
    \centering
    \includegraphics[width=0.9\linewidth]{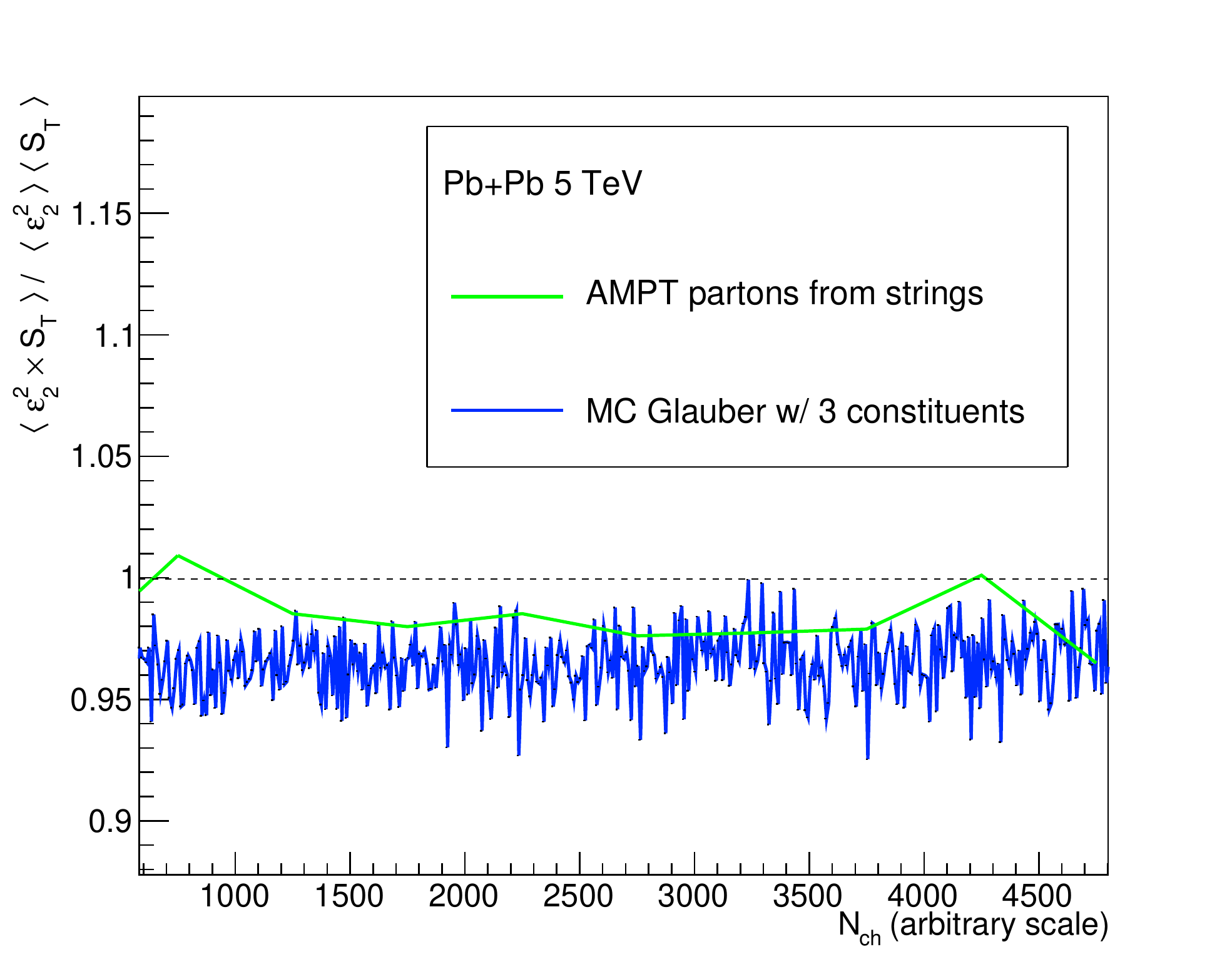}
    \caption{Comparison of the correlation between overlap area and eccentricity in \pbpb collisions  at 5 TeV from \ampt and Monte Carlo Glauber.   The $x$-axis highlights the larger \nch range.}
    \label{fig:ampt_geom_pbpb_largerange}
\end{figure}

Finally, for the \pbpb case, the geometry correlation is opposite in  the very low \nch region between \ampt and Monte Carlo Glauber, where again \ampt may have the wrong geometry as indicated by the lack of sign change that is seen in data -- see Figure~\ref{fig:ampt_pbpb}.   However, for larger \nch as shown in Figure~\ref{fig:ampt_geom_pbpb_largerange}, the correlation in both \ampt and Monte Carlo Glauber is negative and thus would predict a positive $\rho$ correlator -- exactly as seen in Figure~\ref{fig:ampt_pbpb} at $\nch > 250$.   It is notable that the correlation is weaker, i.e. closer to one, in \ampt, though it over-predicts $\rho$ by almost a factor of two.

\section{Summary}
\label{sec:conclusion}

In summary, we have carried out a study of elliptic flow -- transverse momentum correlations within the context of pure non-flow \pythia and \angantyr models and the combination of flow and non-flow \ampt model.    The results indicate that even with  the three-subevent method to  reduce non-flow,  substantial non-zero $\rho$ correlations persist for small collision systems and peripheral event classes in large collision systems.   The \pythia and \angantyr correlations are comparable in size to  predicted glasma correlations.    It is striking that \ampt calculations have the opposite sign to measured ATLAS data in \ppb and peripheral \pbpb collisions, which may indicate that the string geometry, combined with  string melting, has an incorrect modeling of the initial geometry.   Finally, the pure glasma results for the $\rho$ correlator are interesting,  and the simple color domain explanation appears incorrect, and thus  more study is warranted as explanations can be as important as the result itself.

\section{Acknowledgements}

We gratefully acknowledge useful discussions with Giuliano Giacalone, Jiangyong Jia, Blair Seidlitz, Bjoern Schenke, Christian Bierlich, and Ross Snyder.   We also  want to remember Jack Sandweiss, who  was an inspiration to  so many of us and would want us to  continue to ``take nature  to the mat'' and uncover it's secrets.
JLN acknowledges support from the U.S. Department of Energy, Office of Science, Office of Nuclear Physics under Contract No. DE-FG02-00ER41152.   
SHL acknowledges support from the National Research Foundation of Korea (NRF) grant
funded by the Korea government (MSIT) under Contract No. 2020R1C1C1004985.

\clearpage
\pagebreak

\section{Appendix A}
\label{appendixA}

\begin{figure*}[h!]
    \centering
    \includegraphics[width=1.0\linewidth]{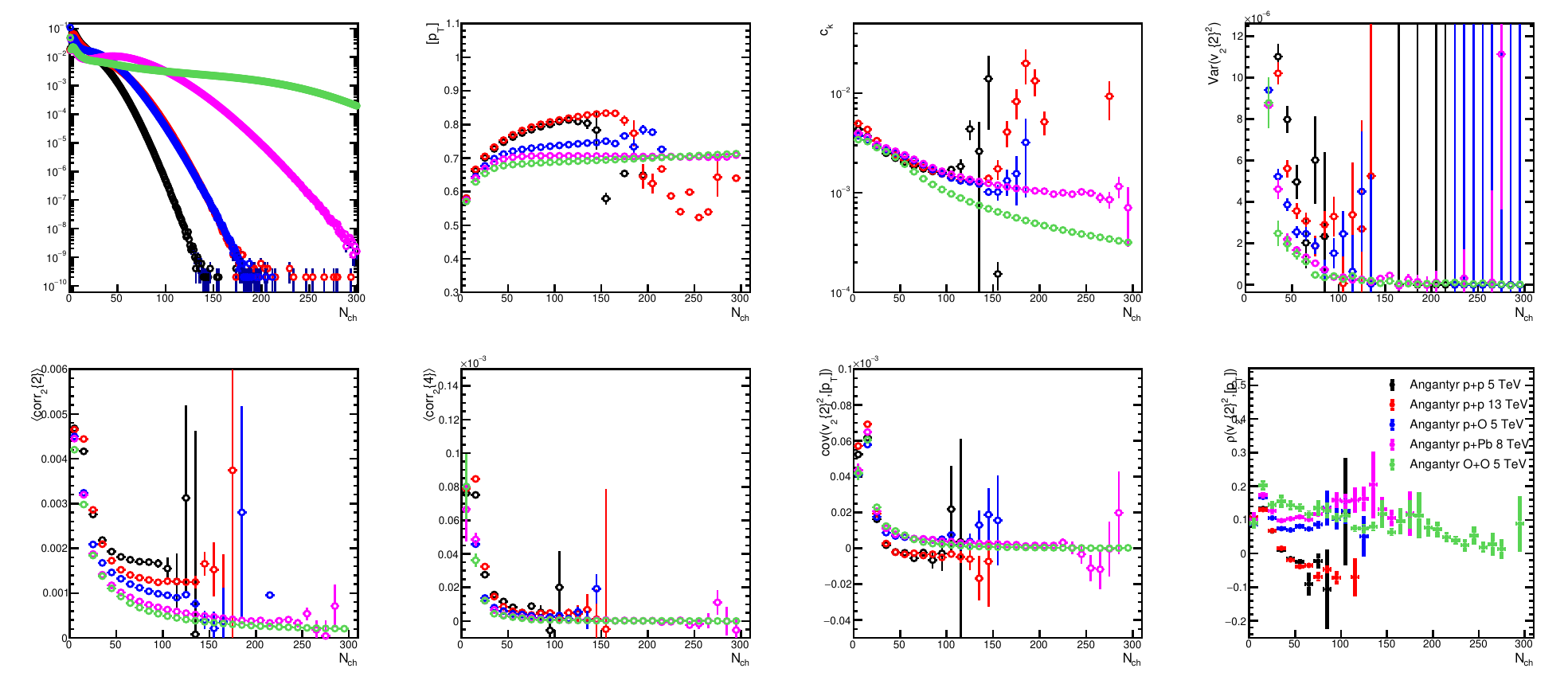}
    \caption{Results from the \pythia model for \pp at 5 and 13 TeV, and the \angantyr model for \po and \oo at 5 TeV and \ppb at 8 TeV are shown.    The quantities in the subpanels are defined in the text.}
    \label{fig:pythiasmall8}
\end{figure*}

For completeness we include the full  set of calculated quantities in the model calculations.   Shown in Figure~\ref{fig:pythiasmall8} are the \pythia and \angantyr results from upper left to upper right and then lower left to lower right: (i) the minimum bias \nch distribution with the integral normalize to one,  (ii) the event $\langle p_{T} \rangle$ as a function of \nch, (iii) the coefficient $c_k$, (iv) the variance of $v_{2}^{2}$, (v) the average two-particle cumulant, (vi) the average four-particle cumulant, (vii) the covariance, and finally (viii) the $\rho$ correlation all as a function of \nch.   We highlight that the range of \nch shown here is meant to  highlight small systems, and thus the \oo distribution extends beyond the range of the figure. Also, at some large \nch at the very tail of the distribution for each collision systems, the statistical uncertainties get large.   In  the $\rho$ distributions, we have eliminated points with uncertainties larger than a set value for clarity of comparison between systems. Note that results from \pythia and \angantyr for \pp at 13 TeV are compared, and no significant difference is observed.

\begin{figure*}[h!]
    \centering
    \includegraphics[width=1.0\linewidth]{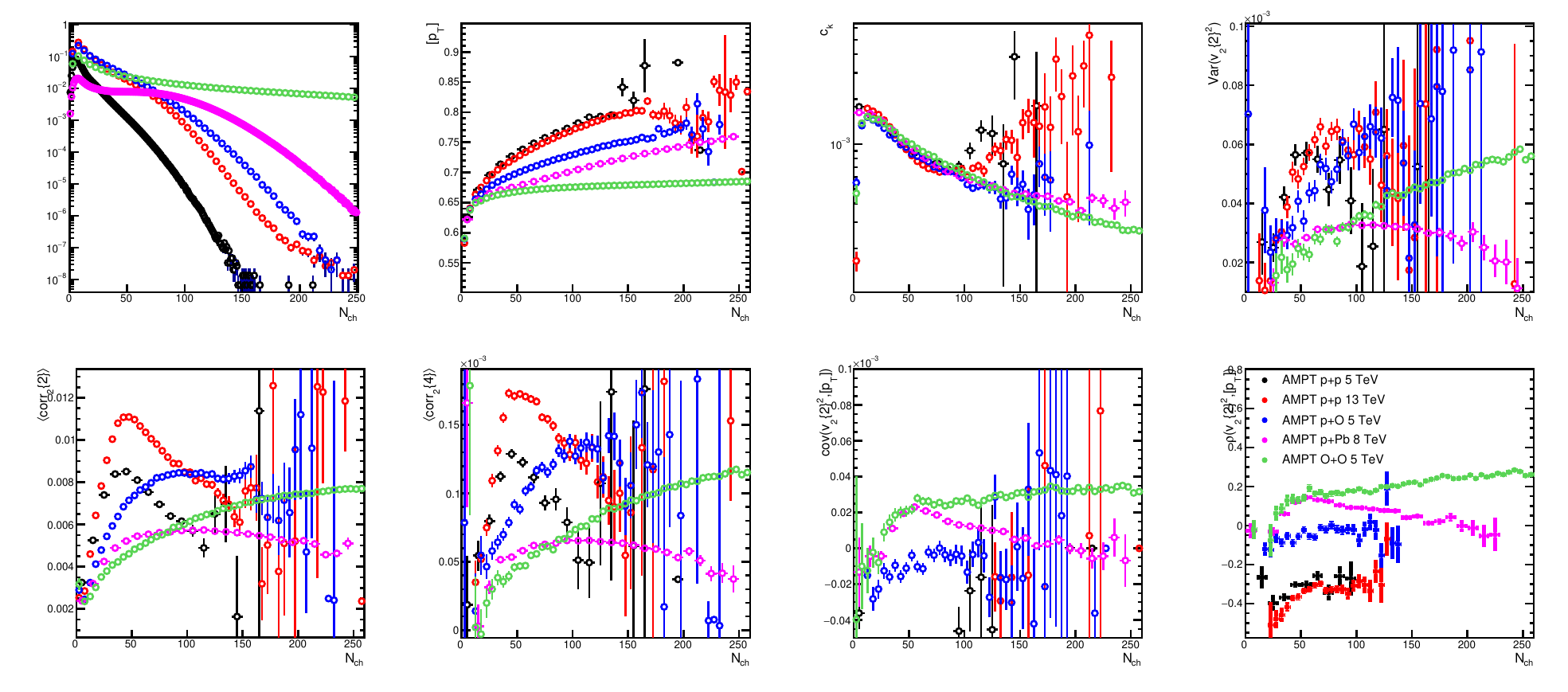}
    \caption{Results from the \ampt model for \pp at 5 and 13 TeV, \po and \oo at 5 TeV,  and \ppb at 8 TeV are shown.    The quantities in the subpanels are defined in the text.}
    \label{fig:amptsmall8}
\end{figure*}

\begin{figure*}[h!]
    \centering
    \includegraphics[width=1.0\linewidth]{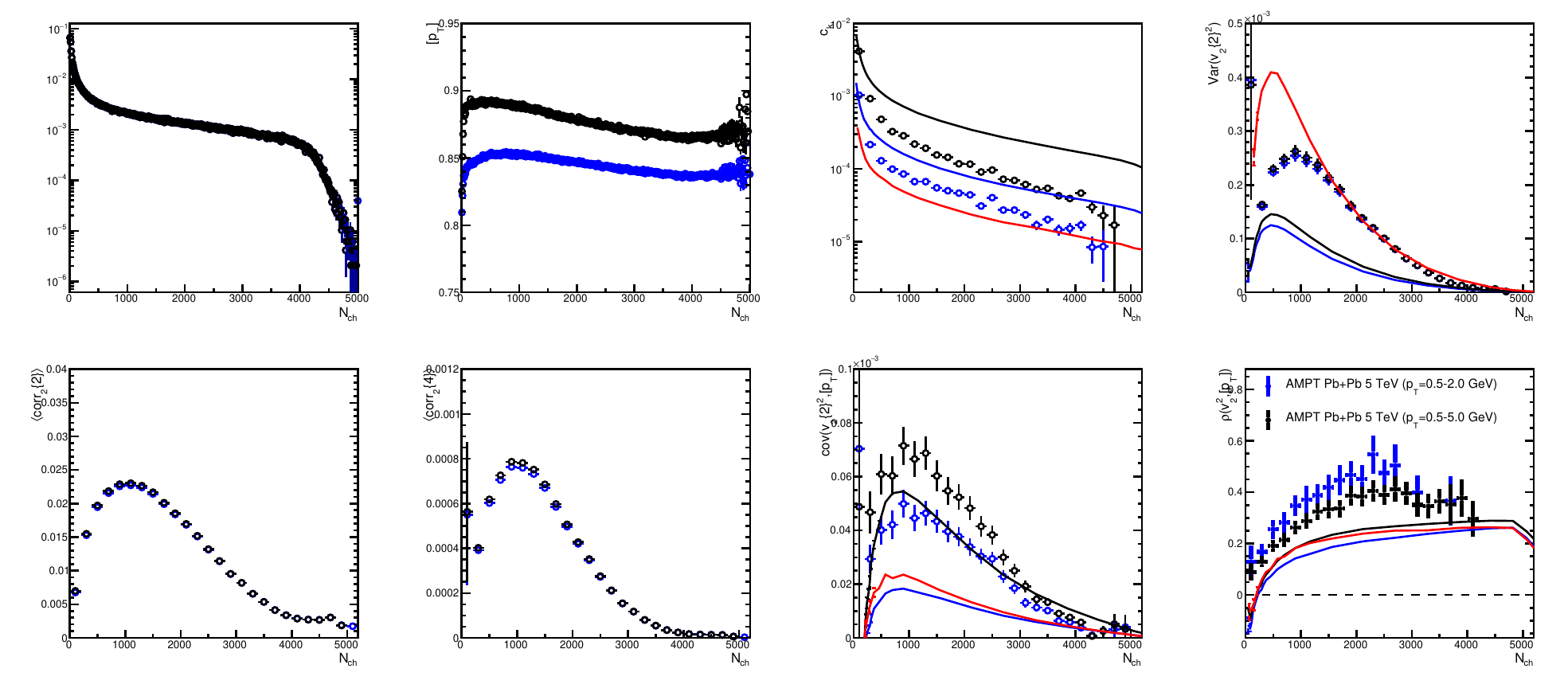}
    \caption{Results from the \ampt model for \pbpb at 5 TeV are shown as circular markers.    The quantities in the subpanels are defined in the text.   Also shown are the ATLAS results for the quantities in the four right subpanels plotted as lines.   The colored lines are for different $p_{T}$ selections -- $0.5 < p_{T} < 2.0$~(blue), $0.5 < p_{T} < 5.0$~GeV (black), and $1.0 < p_{T} < 2.0$~GeV (red).}
    \label{fig:amptbig8withdata}
\end{figure*}

\begin{figure*}[h!]
    \centering
    \includegraphics[width=1.0\linewidth]{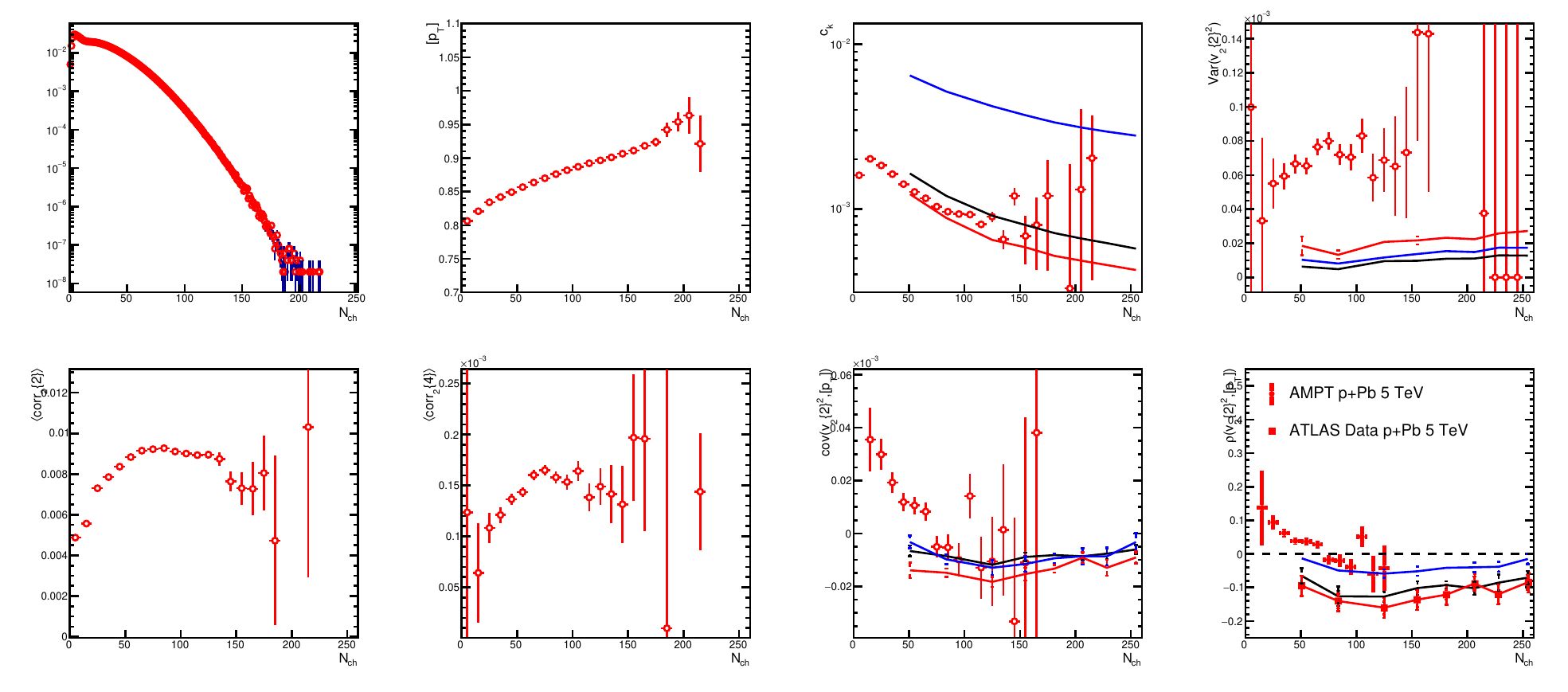}
    \caption{Results from the \ampt model for \ppb at 5 TeV are shown as circular markers. The quantities in the subpanels are defined in the text.   Also shown are the ATLAS results for the quantities in the four right subpanels plotted as lines.   The colored lines are for different $p_{T}$ selections -- $0.3 < p_{T} < 5.0$~(blue), $0.3 < p_{T} < 2.0$~GeV (black), and $0.5 < p_{T} < 2.0$~GeV (red).}
    \label{fig:amptsmall8withdata}
\end{figure*}

Shown in Figure~\ref{fig:amptsmall8} are the same quantities for small systems from \ampt.
Lastly, in Figures~\ref{fig:amptbig8withdata} and ~\ref{fig:amptsmall8withdata} are the eight panels for \pbpb and \ppb at 5 TeV compared with ATLAS results~\cite{Aad:2019fgl}.

\begin{figure*}[h!]
    \centering
    \includegraphics[width=1.0\linewidth]{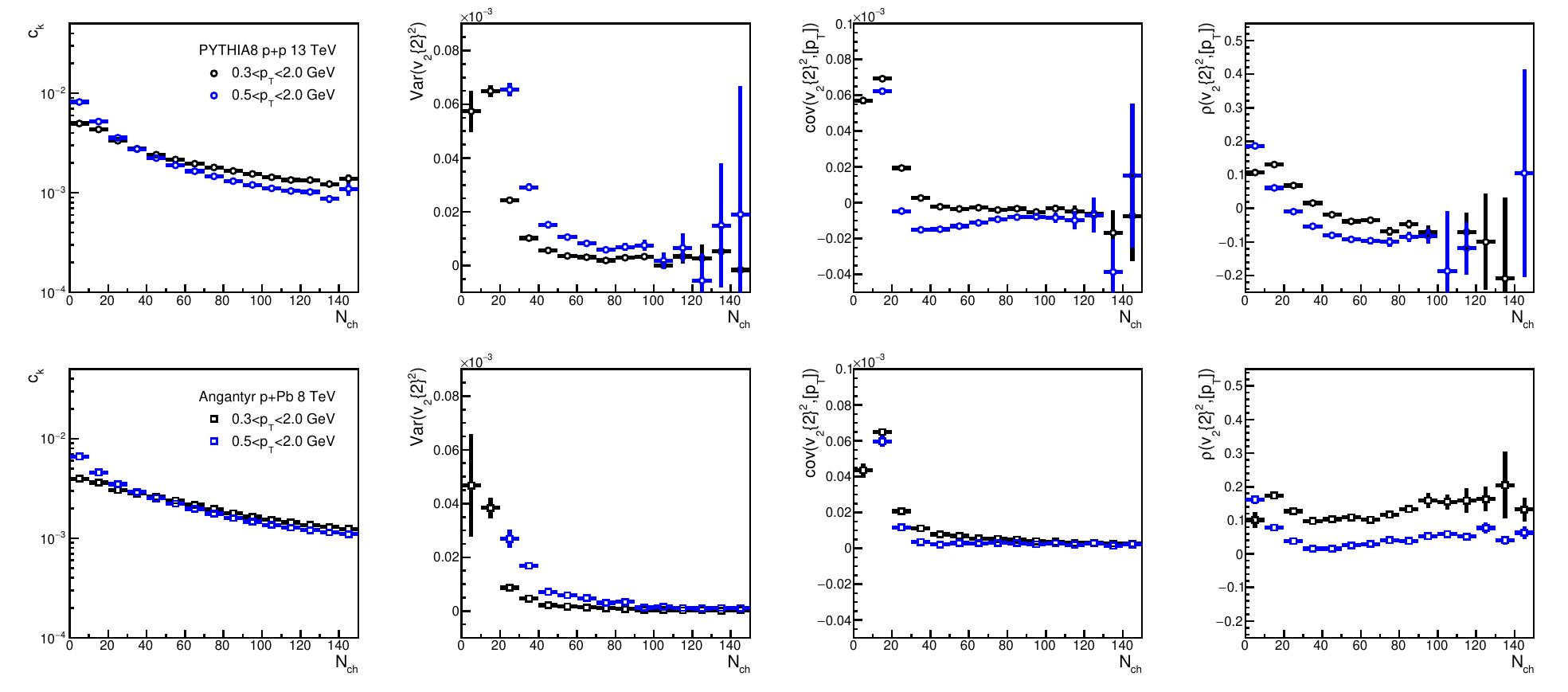}
    \caption{Results from the \pythia model for \pp at 13 TeV and the \angantyr model for \ppb at 8 TeV are shown.    The quantities in the subpanels are defined in the text.   Results for two \pt selections, $0.3 < p_{T} < 2.0~\mathrm{GeV}$ and $0.5 < p_{T} < 2.0~\mathrm{GeV}$, are presented.}
    \label{fig:pythiaptrange}
\end{figure*}

Figure~\ref{fig:pythiaptrange} shows results of the coefficient $c_k$, the variance of $v_{2}^{2}$, the covariance $\mathrm{cov}(v_{2}^{2},[p_T])$, and the $\rho$ correlation for two \pt ranges in \pp at 13 TeV and \ppb at 8 TeV.
The $\rho$ values in $0.5 < p_{T} < 2.0~\mathrm{GeV}$ are lower than those in $0.3 < p_{T} < 2.0~\mathrm{GeV}$ where the non-flow contribution is expected to be smaller.
The $\rho$ correlations at low multiplicity are positive in both \pt ranges and collision systems, but the sign change has been reported in \ppb from \hijing~\cite{zhang2021nonflow}.

\clearpage

\section{Appendix B}
\label{AppendixB}

Here we briefly explore the  relationship between area, eccentricity, and saturation scale within the \ipjazma  framework~\cite{Nagle:2018ybc}.    The \ipjazma calculation includes the IP-Sat features of treating each nucleon, or in  this case each constituent quark, as a two-dimensional Gaussian distribution for the saturation scale $Q_{s}^{2}$.    The saturation scale is allowed to fluctuate via a Gaussian parameter $\sigma = 0.5$ on a log-normal distribution -- thus the fluctuations are quite large and have a high-side tail.    These features mimic the \ipglasma calculations and then the energy deposit is calculated on a grid in  each cell as the product of the summed $Q_{s}^{2}(x,y)$ in the projectile and the summed $Q_{s}^{2}(x,y)$ in the target.   This calculation mimics the dense-dense limit of saturation calculations and reproduces the \ipglasma geometry  with good accuracy~\cite{Snyder:2020rdy}.     We  can also use \ipjazma to mimic the dilute-dense limit where the target contribution is treated logarithmically -- see Ref.~\cite{Nagle:2018ybc} for  details.

We simulate \ppb collisions  at 5 TeV and categorize each  event via the overlap area, the eccentricity, and the saturation scale of the projectile over the overlap region.    For the overlap region, to avoid over-counting the area in cases where there may be so-called ``islands'' of energy deposit (i.e. where different regions are disconnected) we simply count cells above a minimum energy rather than use the pocket equation $S_T = \pi \sqrt{\sigma_{x}^{2}\sigma_{y}^{2}-\sigma_{xy}^{2}}$.
Similarly for the saturation scale of the projectile, we simply average the $Q_{s}^{2}$ value over the cells included in the area calculation.   The absolute values will thus have a sensitivity to this minimum energy,  but for the correlations the relative values should be mostly insensitive.

\begin{figure*}[h!]
    \centering
    \includegraphics[width=0.9\linewidth]{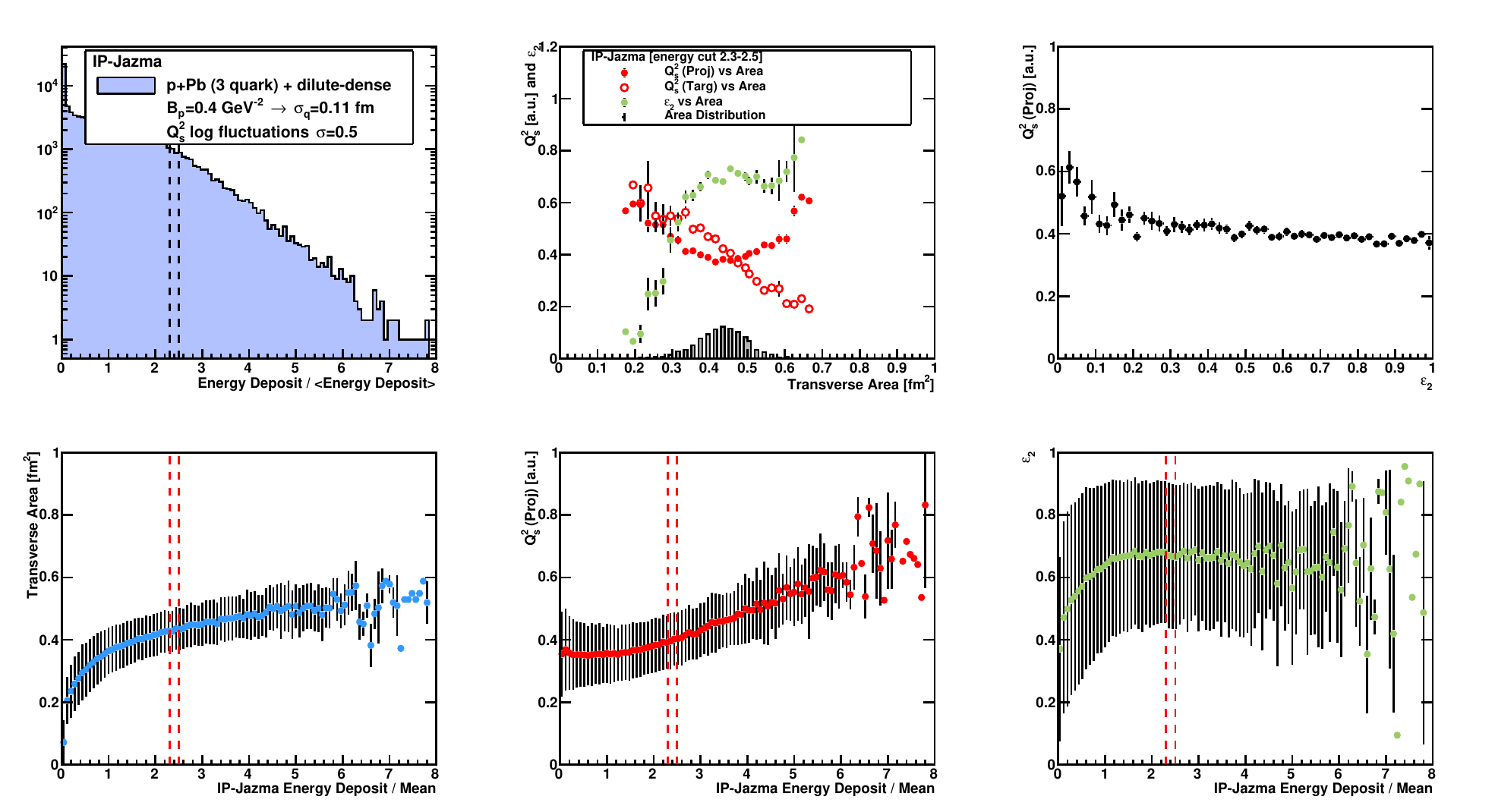}
    \caption{Results from the \ipjazma model for \ppb collision geometry in the dilute-dense case.   Details on the various panels  is given in the main text.}
    \label{fig:jazmadilutedense}
\end{figure*}

\begin{figure*}[h!]
    \centering
    \includegraphics[width=0.9\linewidth]{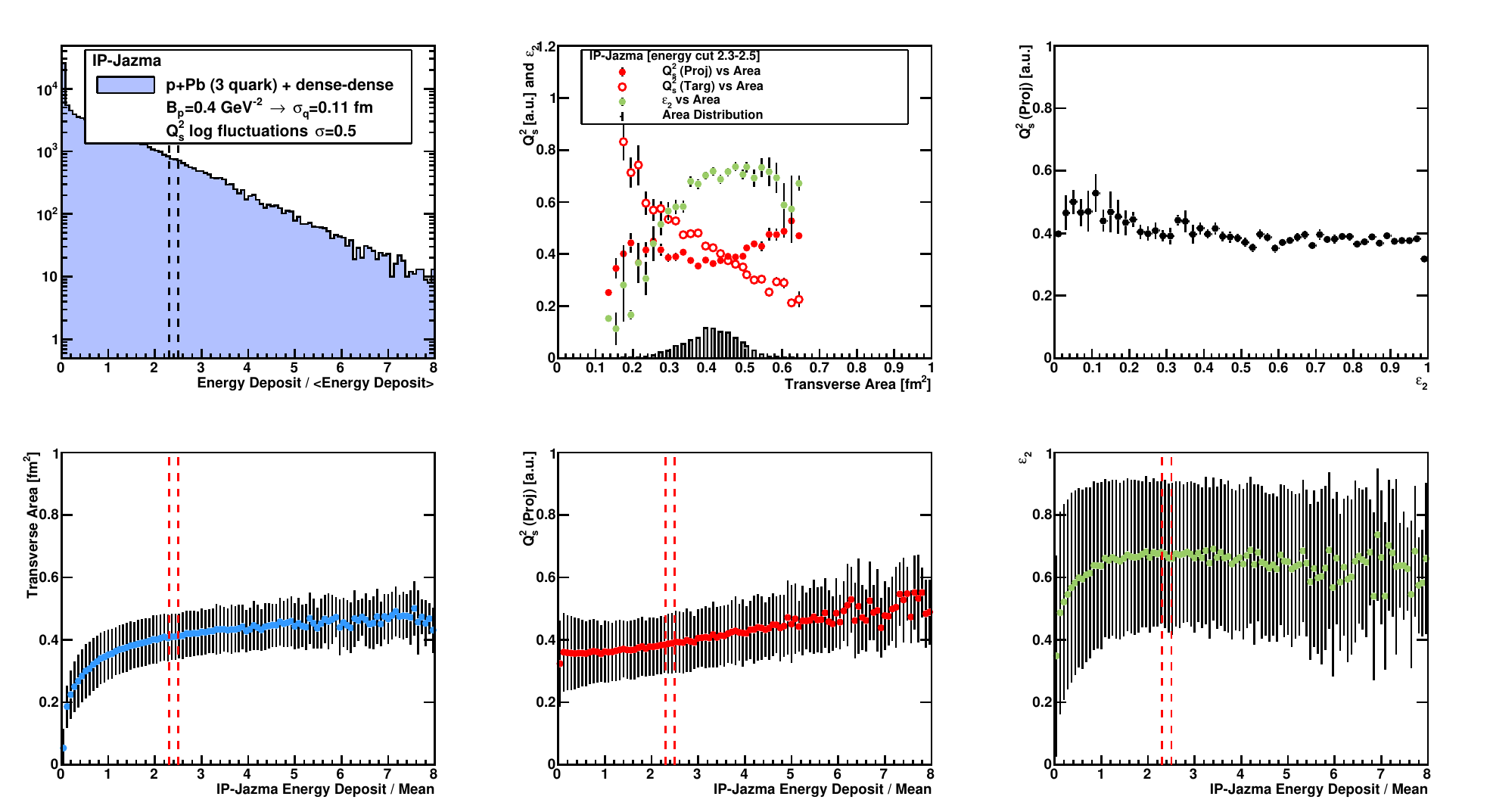}
    \caption{Results from the \ipjazma model for \ppb collision geometry in the dense-dense case.   Details on the various panels  is given in the main text.}
    \label{fig:jazmadensedense}
\end{figure*}

Figure~\ref{fig:jazmadilutedense} shows the dilute-dense case for \ppb collisions.    The upper left panel shows the energy deposit distribution normalized to the mean for minimum bias collisions.   The lower  panels from left to right show the overlap area in the transverse plane, the average  projectile, and separately target, $Q_{s}^{2}$ over the overlap area, and the eccentricity as a function of the energy deposit normalized to the mean.
The points are the mean values while the vertical  lines represent the rms of the distribution about the mean.   We then select a fixed energy deposit bin, as  shown by the vertical dashed lines.    The upper middle panel shows the variation in $Q_{s}^{2}$ and $\varepsilon_{2}$ with transverse area for this fixed energy selection.   Lastly, in the upper right, again in this fixed  energy bin, the correlation of the projectile $Q_{s}^{2}$ and $\varepsilon_{2}$ is shown.    The same results but in the  dense-dense case are shown in Figure~\ref{fig:jazmadensedense}.

One simple remark is that the projectile saturation scale does not increase with decreasing area as conjectured in Ref.~\cite{Schenke:2021mxx}, and rather  has a more complicated dependence.      The very large fluctuations in $Q_{s}^{2}$ added by hand for each constituent  quark play a key role here.    At fixed energy deposit, to have a small  area, it is  most likely that one quark has fluctuated  to  a very large $Q_{s}^{2}$,  which would then explain the anti-correlation of $Q_{s}^{2}$ projectile  and area, for cases where the area is less than 0.4~fm$^{2}$.    It also then explains the very large drop off  in eccentricity, since the  single  quark  distribution is circularly  symmetric.     The behaviour for area greater than 0.4~fm$^{2}$ has an increasing $Q_{s}^{2}$ projectile, and since it is at fixed energy deposit, can only be explained by  having a decrease in the $Q_{s}^{2}$ target in the region of overlap -- also shown in the upper middle panel of Figure~\ref{fig:jazmadilutedense}.   It may seem surprising that the $Q_{s}^{2}$ in the proton projectile and Pb target are so similar, but  this is a consequence of higher multiplicity events arising from large $Q_{s}^{2}$ fluctuations in the projectile -- see Ref.~\cite{Nagle:2018ybc} for example.   Arguments about the expectations ignoring the very large spatial and $Q_{s}^{2}$ fluctuations in the target need to be revisited.   

In  both  the  dilute-dense and dense-dense cases, the relationship between $Q_{s}^{2}$ projectile  and eccentricity is a modest anti-correlation, almost uncorrelated in the  dense-dense case.    Thus, from this simple viewpoint, one might expect a negative $\rho$  correlator.    In contrast, the \ipglasma yields a positive  $\rho$~\cite{Giacalone:2020byk}.    Thus, further work is warranted to understand the source of this positive correlation.

\clearpage

\bibliography{main}

\end{document}